\documentclass[preprint,12pt]{revtex4}

\usepackage[brazil, english]{babel}
\usepackage[utf8]{inputenc}
\usepackage{graphicx}
\usepackage{epsfig}
\usepackage{amssymb}
\usepackage{amsmath} 

\usepackage[mathscr]{euscript}
\usepackage{mathrsfs}
\usepackage{array}
\usepackage{slashed}
\usepackage{dsfont}
\usepackage[normalem]{ulem}

\graphicspath{%
    {converted_graphics/}% inserted by PCTeX
    {/}% inserted by PCTeX
}
\begin{document}

%\date{}
\title{Quark-antiquark potential from a deformed AdS/QCD}
\author{Rodrigo C. L. Bruni$^{1,}$}
\email[Eletronic address:]{bruni.r.c.l@gmail.com}
\author{Eduardo Folco Capossoli$^{2,3,}$}
\email[Eletronic address:]{educapossoli@if.ufrj.br}
\author{Henrique Boschi-Filho$^{3,}$}
\email[Eletronic address: ]{boschi@if.ufrj.br}  
\affiliation{$^1$Departamento de F\'{\i}sica Te\'orica, Universidade do Estado do Rio de Janeiro, 20.550-900 - Rio de Janeiro-RJ - Brazil \\ 
 $^2$Departamento de F\'{\i}sica and Mestrado Profissional em Práticas da Educação Básica (MPPEB), 
 Col\'egio Pedro II, 20.921-903 - Rio de Janeiro-RJ - Brazil\\
 $^3$Instituto de F\'{\i}sica, Universidade Federal do Rio de Janeiro, 21.941-972 - Rio de Janeiro-RJ - Brazil}

\begin{abstract}
In this work we calculate the static limit of the energy for a quark-antiquark pair from the Nambu-Goto action using a holographic approach with a deformed AdS space, with warp factor $\exp\{(\kappa z)^n/n\}$. 
 From this energy we derive the Cornell potential for the quark-antiquark interaction. We also find a range of values for our parameters which fits exactly the Cornell potential parameters.  In particular, setting the zero energy of the Cornell potential at 0.33 fermi, we find that $\kappa=0.56$ GeV and $n=1.3$. 
\end{abstract}

%\pacs{11.25.Wx, 11.25.Tq, 12.38.Aw, 12.39.Mk}

\maketitle

%\begin{multicols}{2}
\section{Introduction}

\noindent

The quark-antiquark potential has been a very useful tool for the investigation of strong interactions and quark confinement. This potential can be used, for example, to analyse the transition between the confined and deconfined phases of matter (see for instance \cite{Gattringer:2010zz}). 

Recently, efforts have been made to obtain the  quark-antiquark potential \cite{Maldacena:1998im, Kinar:1998vq, Boschi-Filho:2006ssg, Andreev:2006ct,White:2007tu, Mia:2010tc, Mia:2010zu, Li:2013oda, Dudal:2017max} using the well known AdS-CFT correspondence. For another approach using effective string theory see for instance \cite{Brambilla:2014eaa}. This correspondence was originally formulated as a mapping of correlation functions of a superconformal $\mathscr{N} = 4$ Yang-Mills theory defined on the boundary of the AdS space and a string theory living in its bulk. It works in such a way that a strongly coupled regime on the boundary theory is mapped into a weakly coupled one in the bulk \cite{Maldacena:1997re,Gubser:1998bc,Witten:1998qj,Witten:1998zw,Aharony:1999ti}.

However, since the original formulation of the correspondence is based on a conformal field theory, which has no characteristic scale, the confining behaviour of the potential is not contemplated once confinement implies a typical length scale. 

In order to describe both the confining and non-confining behaviours, it becomes necessary to break the conformal invariance of the theory. There are various ways of doing so but we mention just two of them: the hardwall \cite{Polchinski:2001tt,Polchinski:2002jw,BoschiFilho:2002vd,BoschiFilho:2002ta,BoschiFilho:2005yh,Capossoli:2013kb,Rodrigues:2016cdb} and the softwall \cite{Karch:2006pv,Capossoli:2015ywa,Capossoli:2016kcr, FolcoCapossoli:2016ejd} models which break conformal invariance introducing a cut off in the action. Inspired by \cite{Andreev:2006ct}, here we break the conformal invariance modifying the background metric instead of the bulk action. So  the metric is given by: 
\begin{eqnarray}\label{metric}
d{\rm s}^{2} = g_{mn}dX^{m}dX^{n}=\frac{R^{2}}{z^{2}}\, h(z) \, (dx^{i}dx_{i}+dz^{2}) \,, 
\end{eqnarray} 
where $R$ is the AdS radius, $m,n=0,1,2,3,z$, where $z$ is the holographic coordinate while $x^i$ with $i=0,1,2,3$ represents an Euclidean space in four dimensions. The warp factor that we consider here in this work is given by:
\begin{equation}\label{h(z)}
h(z) = \exp\left\{ \frac 1n {(\kappa z)^{n}}\right\}\,,
\end{equation}
in which $\kappa$ has dimensions of inverse length and $n$ is a dimensionless number. We will keep these constants arbitrary until sec. \ref{phen},  where we relate our results to phenomenology of the quark antiquark potential. Note that, if we restrict $n=2$ we reobtain  the results of \cite{Andreev:2006ct}.

The main goal of this work is to calculate the energy configuration for a quark-antiquark pair from the Nambu-Goto action using a holographic approach within the deformed metric Eq. \eqref{metric} with the warp factor given by Eq. \eqref{h(z)}. From this energy we will obtain the Cornell potential \cite{Eichten:1978tg,Eichten:1979ms,Eichten:1995ch,Eichten:1994gt} (for excellent reviews of the Cornell potential see \cite{Bali:2000gf, Brambilla:2004jw}): 
\begin{equation}\label{cornell}
V(\ell)= -\frac{\xi}{\ell} + \frac{\ell}{a^{2}} + C \,\, , 
\end{equation}
and also find a range of values for the parameters $\kappa$ and $n$ which describe $h(z)$ in order to fit this potential. 

This work is organised as follows. In Section \ref{WL},  using the warp factor $\exp\{(\kappa z)^n/n\}$, we compute the separation and  the energy of the quark-antiquark pair using the Wilson loop from the AdS/CFT correspondence. In section \ref{phen}, we discuss the matching of our parameters $\kappa$ and $n$ to fit the Cornell potential. Finally, in section \ref{concl}, we present our comments and conclusions. We also include an Appendix where we give some details of the calculation of the energy and the separation distance of the string. 

%%%%%%%%%%%%%%%%%%%%%%%%%%%%%%%%%%%%%%%%%%%%%%%%%
%%%%%%%%%%%%%%%%%%%%%%%%%%%%%%%%%%%%%%%%%%%%%%%%%

\section{The Wilson loop and the quark potential}\label{WL}

The starting point of our calculations involves the Wilson Loop. For convenience we choose one circuit corresponding to a rectangular spacetime loop with temporal extension $T$ and spatial extension $\ell$ in the association with the area of the string worldsheet that lives in the AdS space, whose boundary is just the flat spacetime in 4 dimensions where the loop is defined \cite{Maldacena:1998im, Kinar:1998vq}. 

So, following this prescription, we just have to calculate the Nambu-Goto action of a string with the endpoints (identified as the quark and antiquark) fixed at  $z=0$, assuming a ``U-shape'' equilibrium configuration in the bulk of deformed AdS.	

Assuming also that the string configuration is, by hypothesis, static {\it i.e.} it moves in the interior of the deformed AdS without change in its shape, one can show that the interquark separation and energy for the type of metric \eqref{metric} are respectively given by (see the Appendix for details):
\begin{eqnarray}\label{e}
\ell &=&  2\int_{0}^{z_0}\frac{z^{2}}{z_{0}^{2}}\frac{h(z_{0})}{h(z)} \frac{1}{\sqrt{1 - \left( \frac{h(z_{0})}{h(z)} \right)^{2}\frac{z^{4}}{z_{0}^{4}}}}dz \,,
\\ 
E &=& \frac{1}{\pi\alpha'} \int_{0}^{z_0}\frac{ R^2}{z^{2}}\, h(z)\, \frac{1}{\sqrt{1 - \left( \frac{h(z_{0})}{h(z)} \right)^{2}\frac{z^{4}}{z_{0}^{4}}}}dz \,.  \label{l}
\end{eqnarray}
 Note that $z_0$ is the minimum of the $z$ coordinate and corresponds to the bottom of the U-shape curve. 

The form of \eqref{e} and \eqref{l} is very convenient because it makes explicit that the expressions of energy and separation distance depend only on the warp factor chosen for the metric and the value of $z_0$. 

It is useful to rewrite the integrals \eqref{e} and \eqref{l} in terms of a dimensionless variable. If we define $v := \frac{z}{z_{0}}$, the integrals become: 
\begin{eqnarray}
\ell &=& 2 z_{0}\int_{0}^{1}\frac{h(1)}{h(v)}v^{2} \left[ 1 - \left( \frac{h(1)}{h(v)} \right)^{2}v^{4} \right]^{-\frac{1}{2}} dv  . \label{eqn47} \\  
E &=& \frac{R^{2}}{\pi\alpha'}\frac{1}{z_{0}}\int_{0}^{1}h(v)v^{-2} \left[ 1 - \left( \frac{h(1)}{h(v)} \right)^{2}v^{4} \right]^{-\frac{1}{2}} dv  , \label{eqn48}
\end{eqnarray}
which makes explicit the dimensions of $\ell$ and $E$ since the integrals are now dimensionless, and where we identify $h(v)\equiv h(z)$. Note also that the ratio $R^2/\pi\alpha'$ is dimensionless. 

Now we introduce the dimensionless parameter $\lambda := (\kappa z_{0})^{n}$  such that the equations \eqref{eqn47} and \eqref{eqn48} become:
\begin{eqnarray}
\ell = 2 \frac{\lambda^{\frac{1}{n}}}{\kappa}\int_{0}^{1}v^{2} e^{\frac{\lambda}{n}(1-v^{n})}\left( 1 - e^{\frac{2\lambda}{n}(1-v^{n})} v^{4} \right)^{-\frac{1}{2}} dv  , \label{lf}  \\ 
E = \frac{R^{2}}{\pi\alpha'}\frac{\kappa}{\lambda^{\frac{1}{n}}}\int_{0}^{1}e^{\frac{\lambda}{n}v^{n}}v^{-2}\left( 1 - e^{\frac{2\lambda}{n}(1-v^{n})}v^{4} \right)^{-\frac{1}{2}} dv  , \label{Ef}
\end{eqnarray}
where $\kappa$ has the dimension of energy. 
Let us analyze the above expressions when $\lambda \approx 0$ and $\lambda \approx 2$, which are the interesting physical limits since for $\lambda \to 0$ one has $\ell \to 0$, while for  $\lambda \to 2$ one has $\ell \to \infty$, as we are going to discuss below. 

\subsection{Calculation of $\ell$}

\subsubsection{$\lambda$ close to zero} \label{sub1.1}

If we express the integrand in \eqref{lf} as a power series in $\lambda$ centered at zero, to first order in $\lambda$ and integrate it, we obtain:
\begin{eqnarray}\label{eqn52}
I(\lambda,n) = -\frac{\sqrt{\pi }}{2 n} \left(\frac{\Gamma \left(\frac{3}{4}\right) (\lambda-2 n)}{\Gamma \left(\frac{1}{4}\right)}-\frac{\lambda \Gamma \left(\frac{n+3}{4}\right)}{\Gamma
   \left(\frac{n+1}{4}\right)}\right) ,
\end{eqnarray}
where the above result is valid only if $n>-3$, otherwise the integral does not converge. 

Substituting this result in \eqref{lf} and grouping terms proportional to $\lambda$ one finds:
\begin{eqnarray}\label{eqn53}
\ell = \frac{1}{\rho_0}\frac{\lambda^{\frac{1}{n}}}{\kappa}  \left\lbrace 1 - \frac{\lambda}{2n}  \left[1 - F(n) \pi \rho_0   \right] \right\rbrace + \mathscr{O}(\lambda^{2})  \,; \qquad (\lambda \approx 0).
\end{eqnarray}
where we have defined the dimensionless number $\frac1{\rho_0} := \frac{(2\pi)^{\frac{3}{2}}}{\Gamma^{2} \left( \frac{1}{4} \right )}$ and function $F(n) := \frac{2}{\sqrt{\pi}}\frac{\Gamma \left( \frac{3+n}{4} \right)}{\Gamma \left( \frac{1+n}{4}  \right)}$.

\subsubsection{$\lambda$ close to 2} \label{sub1.2}

If we repeat the procedure of last subsection for $\lambda$ now centered at 2 we will not be able to achieve an analytic expression for the integral. We note however that the integral of \eqref{lf} is dominated by $v \sim 1$. We thus expand the integrand around $v=1$ to first order and integrate it, obtaining:
\begin{eqnarray}\label{eqn54}
I(\lambda, n) &=& \left(\frac{1}{{\sqrt{\lambda (2 \lambda +n-9)+10}}} \right) 
\Big[ -\log \Big(-2 (\lambda - 2) \big[\lambda (2 \lambda +n-9)+10\big]\Big)  \nonumber\\
&+& 2 \log \left(\lambda (2 \lambda + n-9)+\sqrt{[\lambda (2 \lambda +n-11)+14] [\lambda (2 \lambda +n-9)+10]}+10\right) \Big] \nonumber \\ &&
\end{eqnarray}

As the first logarithm of \eqref{eqn54} diverges when $\lambda = 2$ one would expand again around  $\lambda = 2$ up to first order. However, since terms of order $\mathscr{O}(1)$ in the expansion will not contribute to the functional form of the Cornell potential and we are extracting just the leading behavior of \eqref{lf} for $\lambda \approx 2$, we can safely neglect contributions of order $\mathscr{O}(\lambda)$ in the aforementioned expansions, obtaining:
\begin{eqnarray}\label{eqn59}
I(\lambda,n) = -\frac{1}{\sqrt{2n}} \log(2-\lambda) + \mathscr{O}(1)  \,,
\end{eqnarray}
which, due to \eqref{lf} leads to:
\begin{eqnarray}\label{eqn60}
\ell =  \frac{2^{\frac{1}{n}}}{\kappa} \left[  - \sqrt{\frac{2}{n}} \log(2-\lambda) + \mathscr{O}(1) \right]  \,; \qquad (\lambda \approx 2). 
\end{eqnarray}
As mentioned above, the limit $\lambda \to 2$ implies  $\ell \to \infty$.

\subsection{Calculation of the energy}

Before we calculate the integral in eq. \eqref{Ef} let us point out that it diverges as $1/v^{2} $ when $v \to 0$. This becomes clear if one analyzes the series expansion of the integrand in  $\lambda$ close to 0 and 2.

So, we choose the renormalization of \eqref{Ef} as:
\begin{eqnarray}\label{Eren}
E_{\rm Ren.} = \frac{R^{2}}{\pi\alpha'}\frac{\kappa}{\lambda^{\frac{1}{n}}}\left\lbrace -1 + \int_{0}^{1}e^{\frac{\lambda}{n}v^{n}}v^{-2}\left[ \left( 1 - e^{\frac{2\lambda}{n}(1-v^{n})}v^{4} \right)^{-\frac{1}{2}} - 1 \right] dv  \right\rbrace \,\, ,
\end{eqnarray}
such that this energy expression is finite and now we can analyse again the limits of $\lambda$ close to 0 and 2.

\subsubsection{$\lambda$ close to zero} \label{sub2.1}

Expanding the integrand in \eqref{Eren}  with respect to $\lambda$, centered at zero, we find:
\begin{eqnarray}\label{eqn64}
I(\lambda, n) = 1 - \frac{1}{2 \rho_0} +  \left[ \frac{\sqrt{\pi } (n+1) \Gamma \left(\frac{n-1}{4}\right)}{8 n \Gamma \left(\frac{n+1}{4}\right)} - \frac{1}{4 n }\frac{1}{\rho_0}\right] \lambda  ,
\end{eqnarray}

So that, the renormalised energy is:
\begin{eqnarray}\label{Eren0}
E_{\rm Ren.} = - \frac{R^{2}}{\pi\alpha'} \frac{1}{2\rho_0} \frac{\kappa}{\lambda^{\frac{1}{n}}}\left\lbrace 1 + \frac{\lambda}{2n} \left[ 1 - G(n)\pi \rho_0 \right] + \mathscr{O}(\lambda^{2})  \right\rbrace  , 
\end{eqnarray}
where we defined the dimensionless function $G(n) = \frac{(n+1) \Gamma \left(\frac{n-1}{4}\right) }{2 \sqrt{\pi}  \Gamma \left(\frac{n+1}{4}\right)} $. 

Writing the pre factor ${\kappa}/\lambda^{\frac{1}{n}}$ as a function of $\ell$ (c.f. \eqref{eqn53}), substituting in \eqref{Eren0} and keeping only linear terms in $\lambda$ we get:
\begin{eqnarray}\label{eqn67}
E_{\rm Ren.} = \frac{R^{2}}{\pi\alpha'} \, \left\{ - \frac{\xi_{0}}{\ell} + \frac{\lambda}{4 \ell} \left[ \frac{G(n) - F(n)}{\rho_0 n} \right] + \mathscr{O}(\lambda^{2}) \right\} , 
\end{eqnarray}
where we defined the dimensionless number $\xi_{0} :=  {1}/{(2 \rho_0^{2}) } $. Using \eqref{eqn53} we can rewrite $\lambda \approx 0 $ in terms of $\rho_0$ and $\kappa$: 
\begin{eqnarray}\label{eqn68}
\lambda  \approx (\kappa \ell \rho_0)^{n}  \left[ 1 + \frac{\lambda}{2}  \left( 1 - F(n) \pi \rho_0  \right) \right] \,.
\end{eqnarray} 

Substituting this result in  \eqref{eqn67} and keeping in mind that $\lambda \approx 0$ is equivalent to the regime of short distances, one can safely disregard terms proportional to $\ell^{2n-1}$ in comparison with the terms proportional to $\ell^{n-1}$. Then,  we obtain:
\begin{eqnarray}\label{eqn69}
E_{\rm Ren.} = \frac{R^{2}}{\pi\alpha'} \left\lbrace - \frac{\xi_{0}}{\ell} + \sigma_{0}(n)\,  \ell^{n-1} + \mathscr{O}(\ell^{2n-1}) \right\rbrace    , 
\end{eqnarray}
where we defined the function $\sigma_{0}(n) := \frac{1}{4}\kappa^{n} \rho_0^{n-1} \left[ \frac{G(n) - F(n)}{ n} \right]$, with dimensions of (energy)$^n$.

\subsubsection{$\lambda$ close to 2} \label{sub2.2}

In this section we are going to calculate the renormalised energy for $\lambda$ close to 2. 
Repeating the procedure employed in subsection  \ref{sub1.2}, {\it i.e.}, rewriting all the integrand in \eqref{Eren} inside the square root
\begin{eqnarray}\label{eqn70}
I(\lambda , v,  n) =  \left[ v^4 e^{-\frac{2 \lambda v^n}{n}} \left(1-v^4 e^{\frac{2 \lambda \left(1-v^n\right)}{n}}\right)\right]^{-\frac{1}{2}} - v^{-2} \hspace{.1cm} ,
\end{eqnarray}
and expanding this integrand with respect to $v$ centred at 1, to second order we find:
\begin{eqnarray}
I(\lambda , v,  n) = 2 (2-\lambda) (1-v) e^{-\frac{2 \lambda}{n}}- e^{-\frac{2 \lambda}{n}} \left(6 \lambda^2 - \lambda n-23 \lambda + 22\right) (1-v)^2  \nonumber \\ 
+ 3 (1-v)^2 + 2 (1 - v) \hspace{.1cm} . \hspace{2.8cm}\label{eqn72}
\end{eqnarray}
For the above expression to be real, the first two terms must be positive and the last one must be negative which implies, respectively, in $\lambda <2 $ and $\frac{n+23}{12}-\frac{1}{12} \sqrt{n^2+46 n+1}<\lambda<\frac{1}{12} \sqrt{n^2+46 n+1}+\frac{n+23}{12}$.\hspace{.1cm} Now,  integrating  \eqref{eqn72} one has:
\begin{eqnarray}\label{eqn73}
I(\lambda, n) &=& - \, 3 - \frac{\log (4-2 \lambda)}{\sqrt{e^{-\frac{2 \lambda}{n} } \left(-6 \lambda^2+\lambda (n+23)-22\right)}}  \nonumber\\
&& + \frac{2 \log \left[\sqrt{\lambda (-6 \lambda +n+21)-18}\, +\sqrt{\lambda (-6 \lambda+n+23)-22}\, \right] }{\sqrt{e^{-\frac{2 \lambda}{n} } \left[-6 \lambda^2+\lambda (n+23)-22\right]}} . 
\end{eqnarray}

Keeping only terms in lowest order of $\lambda$ and substituting $\lambda = 2$ in the denominator of above expression, we get from Eq. \eqref{Eren}:
\begin{eqnarray}\label{eqn74}
 E_{\rm Ren.} &=& \frac{R^{ 2 }}{\pi\alpha'}\frac{ \kappa}{2^{\frac{1}{n}}}\left\lbrace - \, \frac{e^{\frac{ 2}{n} }\log (2 - \lambda)}{\sqrt{2n}} + \mathscr{O}(1) \right\rbrace \nonumber \\ \nonumber \\ 
&=& \frac{R^{2}}{\pi\alpha'} \left[ \, \sigma(n)\, \ell  + \mathscr{O}(1) \right]  , 
\end{eqnarray} 
where we have used the relation between $\ell$ and $\lambda$ given by Eq. \eqref{eqn60} and defined $\sigma(n) :=  \frac 12 {\kappa^{2}} \left( \frac{e}{2} \right)^{\frac{2}{n}} $. 

\section{Phenomenology}\label{phen}

Summarizing the results of the last section, the renormalised energies \eqref{eqn69} and \eqref{eqn74} in terms of the separation $\ell$ are given by:
\begin{eqnarray}\label{e1}
E_{\rm Ren.}^{\lambda \approx 0} &=& \frac{R^{2}}{\pi\alpha'} \left\lbrace - \frac{\xi_{0}}{\ell} + \sigma_{0}(n)\,  \ell^{n-1} + \mathscr{O}(\ell^{2n-1}) \right\rbrace    ,\\ 
E_{\rm Ren.}^{\lambda \approx 2} &=& \frac{R^{2}}{\pi\alpha'} \left[ \sigma(n) \, \ell  + \mathscr{O}(1) \right]    \, . \label{e2}
\end{eqnarray} 
with
\begin{equation}
\xi_{0} :=  \frac{1}{2 \rho_0^{2}}  \;; \qquad \frac 1 {\rho_0} := \frac{(2\pi)^{\frac{3}{2}}}{\Gamma^{2} \left( \frac{1}{4} \right )} \;; \qquad \sigma(n) :=  \frac 12 {\kappa^{2}} \left( \frac{e}{2} \right)^{\frac{2}{n}} \,. 
\end{equation}
The precise definition of $\sigma_0(n)$, given after Eq. \eqref{eqn69}, will not be needed here since in this section we are going to disregard the term proportional to $\ell^{n-1}$ in comparison with the term of order $\ell^{-1}$, once $n>0$ and in Eq. \eqref{e1}  $\ell \approx 0$. 

Now we are going  to fit the constants of our model with the phenomenological constants of the Cornell potential \eqref{cornell} with $\xi = 0.52$ and $a = 2.34{\rm GeV}^{-1}$ \cite{Eichten:1978tg,Eichten:1979ms,Eichten:1995ch,Eichten:1994gt} (for excellent reviews of the Cornell potential see \cite{Bali:2000gf, Brambilla:2004jw}). 

First of all, we fix the dimensionless ratio $R^{2}/\pi\alpha'$ from the slope of the linear potential at long distances, where the stringy picture is more reliable. Since this regime is equivalent to $\lambda \approx 2$ we compare Eq. \eqref{cornell} with Eq. \eqref{e2}, which leads to the condition $1/a^{2} = R^{2}\sigma(n)/\pi\alpha'$ and therefore:
\begin{eqnarray}\label{eqn76}
\frac{R^{2}}{\pi\alpha'}  = \frac{2}{ (a \kappa )^{2}}\times \left( \frac{2}{e} \right)^{\frac{2}{n}} \;.
\end{eqnarray}

Next, we compare the expression \eqref{e1} with \eqref{cornell}, finding $R^{2}/\pi\alpha'  = \xi/ \xi_{0}$, so that eliminating the ratio $R^{2}/\pi\alpha'$, one obtains
\begin{eqnarray}\label{constraint}
\xi = \frac{1}{(2.34 \kappa )^{2}} \times \left( \frac{2}{e} \right)^{\frac{2}{n}}\frac{1}{ \rho_0^{2} }  \, .
\end{eqnarray}

The above equation can be solved graphically for given values of $\kappa$: we present some of these solutions in Figure 1, for the interval $( \, 0.55 \le \kappa \le 0.70 \, ) \, {\rm GeV}$.  

% It is important to highlight that we can always choose a value of $\kappa$ and a corresponding value of $n$ such that we fit exactly the phenomenological data for the Cornell potential. 
%
\begin{figure}[htb!]
\centering
\includegraphics[scale=.5]{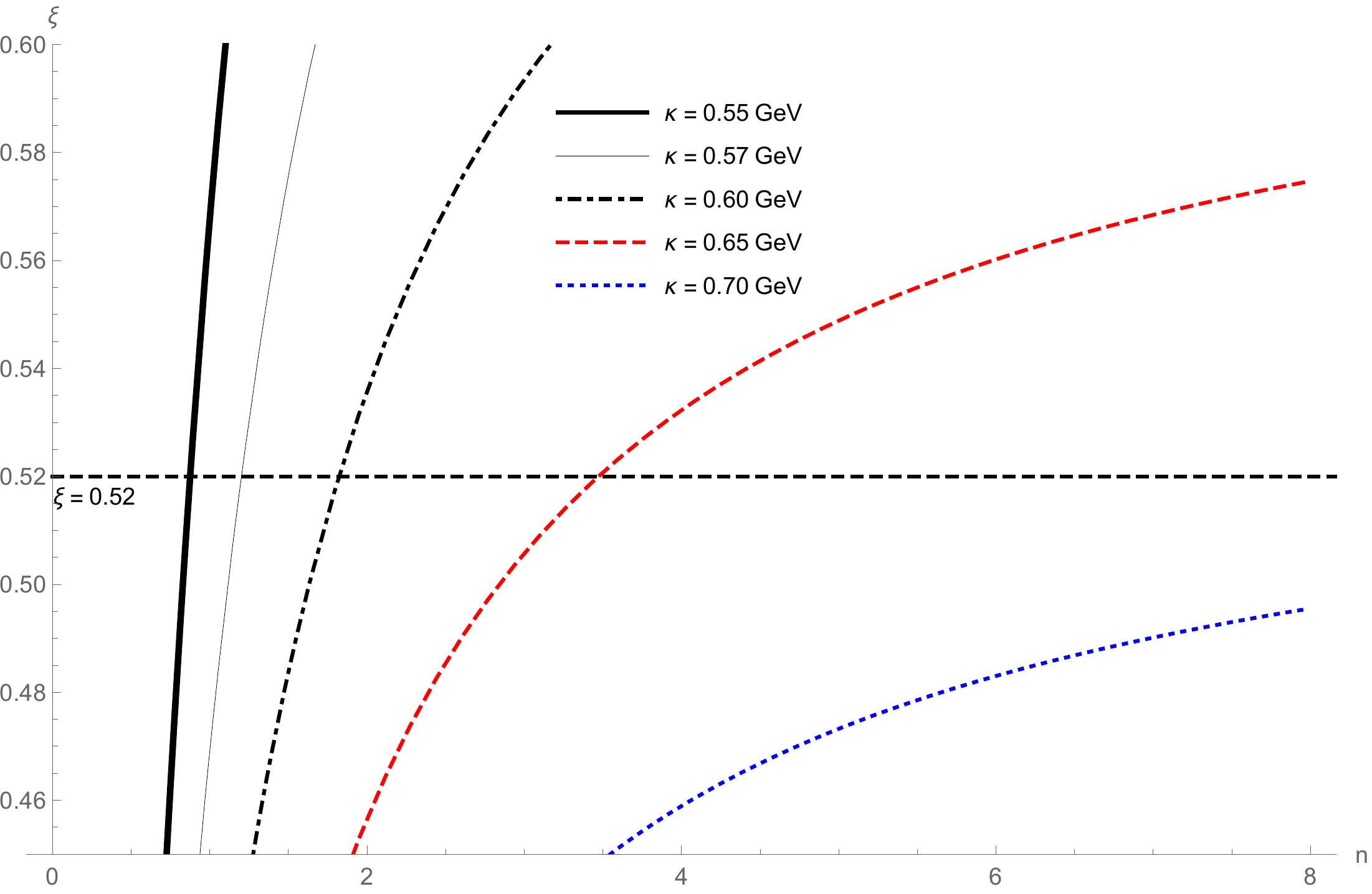}
\caption{Equation \eqref{constraint} solved graphically: The curves are plots of Eq. \eqref{constraint} for some values of $\kappa$ with $a = 2.34 \, {\rm GeV}^{-1}$. The horizontal dashed line represents the phenomenological desired value of the parameter $\xi$, {\it i.e.}, $\xi = 0.52 $ to fit the Cornell potential.}
\label{fit1}
\end{figure}

With the values of parameter $\kappa$ and its corresponding values of $n$ we can investigate the energy associated to the quark-antiquark pair through numerical calculations. In Figure 2 we plot the quark-antiquark potential $E_{Ren.}$ in terms of the quark separation $\ell$, for some values of $\kappa$. 

\begin{figure}[htb!]
\centering
\includegraphics[scale=.4]{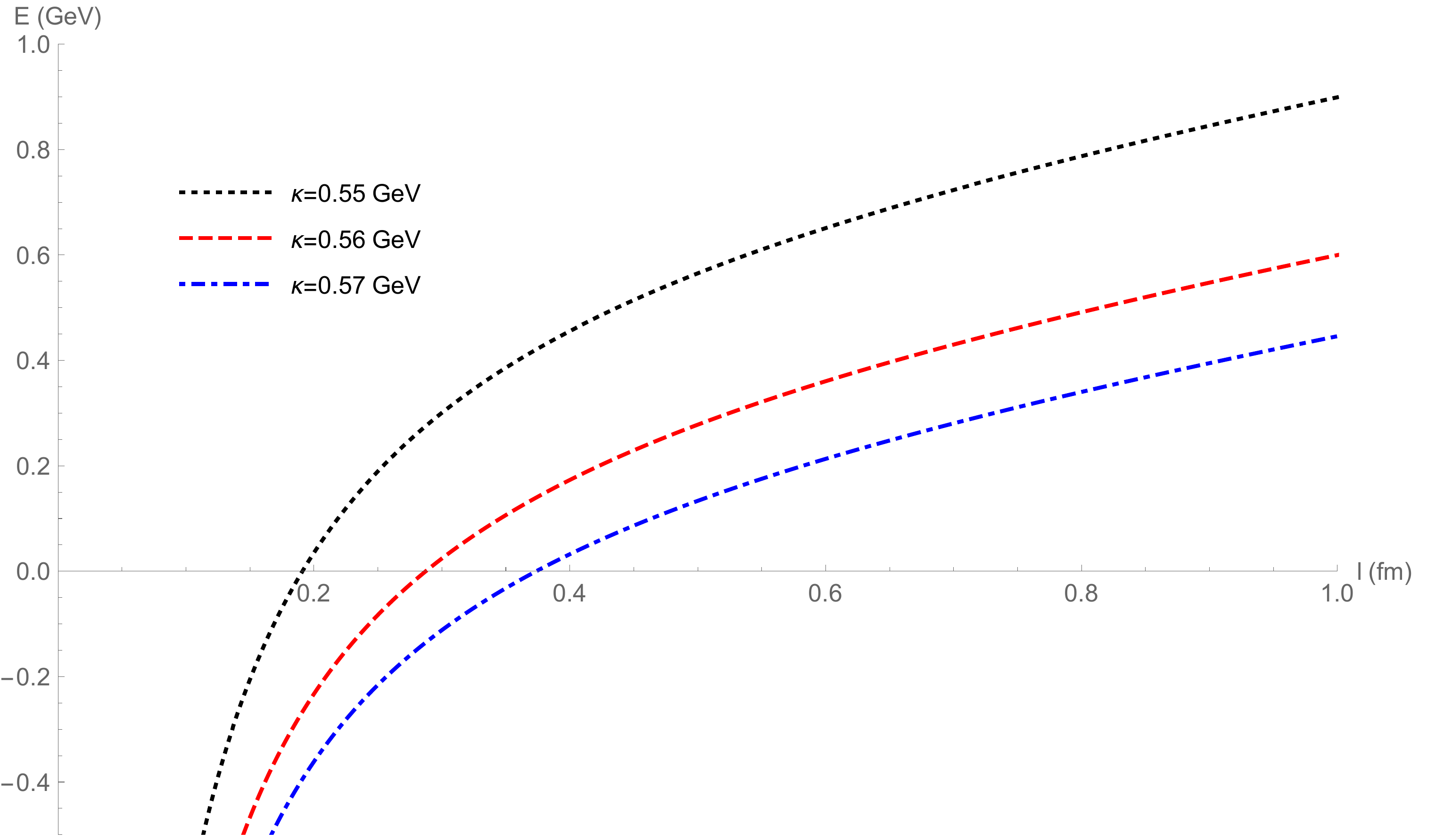}
\caption{$E_{Ren.}$ against $\ell$ obtained directly from Eqs. \eqref{lf} and \eqref{Eren} through numerical integration, for three particular values of $\kappa$: 0.55 ${\rm GeV} $, 0.56 ${\rm GeV} $, 0.57 ${\rm GeV} $  and their respective approximate values of $n$: 1.2, 1.3 , 1.4.  These curves correspond to possible matches with the Cornell potential. The values  $n$ come from  the Figure 1 for each curve corresponding to a given $\kappa$.}
\label{fit2}
\end{figure}

If we fix the constant $C$ in the Cornell potential \eqref{cornell} to be zero, we can obtain a phenomenological constraint such that $V(\ell)=0$ occurs for $\ell\approx 0.33$ fermi. Then, for our warp factor such behaviour is achieved for $\kappa=0.56$ GeV and $n=1.3$, which corresponds approximately to the red dashed line in Figure \ref{fit2}.

Note also that in Figure 2, for the linear confining behaviour all the curves shown present the same slope. This is not a universal property of the deformation we have considered but rather is a choice to fit the Cornell potential parameters.

%\newpage

\section{Concluding remarks} \label{concl}

In this work we have calculated the energy corresponding to a given separation between a quark-antiquark pair from the Nambu-Goto action using a deformed AdS space as a background. The choice of the deformed AdS space is based on the introduction of an exponential factor given by $h(z)=\exp\{(\kappa z)^n/n\}$, Eq. \eqref{h(z)}. We have also shown that this configuration energy has the shape of a Cornell potential. 
In order to fit the Cornell potential parameters we can choose a variety of possibilities for the pair $(\kappa, n)$. In Figure 1, we have shown some of these possibilities and in Figure 2, we presented some profiles for the Cornell potential. 
Note that in Figure 2 we observe the transition from a confining to a non-confining regime around $ \ell \sim 0.3$ fm. 
Specifically, for the choice $\kappa=0.56$ GeV and $n=1.3$, we matched the Cornell potential with the condition $C=0$, as represented by the red dashed line.  

Another interesting feature of our model is the universal non-confining behavior for $\ell \approx 0$, already pointed out by \cite{Maldacena:1998im}. In the context of our model, this universal behavior for short distances is due to the fact that $\ell \approx 0$ is equivalent to $\lambda \approx 0$ ( c.f. section \ref{sub1.1} ) which means that $h(z) \to 1$ and hence we recover the geometry of pure AdS space and therefore we must obtain the non-confining term due to the conformal symmetry of the background space.

Also one can see that our deformation of the AdS, which is a UV deformation, is the fact that it only affects the large distance physics. This modification is encapsulated by the coefficients of the linear term in Eq. \eqref{e2}, which become dependent on the deformation $h(z)$, where {$\kappa$ and $n$, are the parameters that control the deformation. It is interesting that the confining behavior is maintained despite of the choice of $\kappa$ and $n$ which is actually an explicit manifestation of the criterion discussed by \cite{Kinar:1998vq}.

\section*{Conflicts of Interest}

The authors declare that they have no conflicts of interest.

\begin{acknowledgments}
We would like to thank O. Andreev for discussions, and N. Brambilla, K. Dasgupta and S. Mahapatra for useful correspondence. We also would like to thank an anonymous referee for a careful reading of the manuscript and for interesting suggestions. R.C.L.B. is supported by Coordenação de Aperfeiçoamento de Pessoal de Nível Superior (Capes). E.F.C is partially supported by PROPGPEC-Colégio Pedro II and H.B.-F is partially supported by Conselho Nacional de Desenvolvimento Científico e Tecnológico (CNPq) and Capes, Brazilian agencies.
 
\end{acknowledgments}

\section{Appendix}

We start this appendix following \cite{Kinar:1998vq}, defining a metric given by:
\begin{eqnarray}\label{eqn2}
d{\rm s}^{2}= -G_{00}(s)dt^{2}+ G_{x_{||}x_{||}}dx_{||}^{2}+ G_{ss}ds^{2}+G_{x_{T}x_{T}}dx_{T}^{2} \;, 
\end{eqnarray} 
and the Nambu-Goto action:
\begin{equation}\label{eqn3}
S = \frac 1{2\pi\alpha' } \int d\sigma d\tau  \sqrt{\det[\partial_{\alpha}X^{M}\partial_{\beta}X^{N}G_{MN}]} \;.
\end{equation}
\noindent Choosing the gauge $ \sigma=x $ and $\tau =t$ and integrating with respect to $t$, one gets:
\begin{equation}
S = \frac{T}{2\pi\alpha'} \int dx \sqrt{G_{00}(s(x))G_{x_{||}x_{||}}(s(x))+ G_{00}(s(x))G_{ss}(s(x))(\partial_{x} s)^{2}}.
\end{equation}
\noindent where $T$ is the temporal extension of the Wilson loop.

Then, we define \cite{Kinar:1998vq}:
\begin{eqnarray}
f^{2}(s(x)) &=& G_{00}(s(x))G_{x_{||}x_{||}}(s(x)) \label{eqn4} \\
g^{2}(s(x)) &=& G_{00}(s(x))G_{ss}(s(x)).\label{eqn5}
\end{eqnarray}
so that we are left with the integral:
\begin{eqnarray}\label{eqn6}
S = \frac{T}{2\pi\alpha'} \int dx \sqrt{f^{2}(s(x))+g^{2}(s(x))(\partial_{x} s)^{2}} \; .
\end{eqnarray}

Using the differential equation for the geodesic of the string in its equilibrium configuration we get that the separation of the endpoints (or, in our perspective, the quark and antiquark distance) is given by:
\begin{eqnarray}\label{eqn7}
l = \int dx = \int \left(\frac{ds}{dx}\right)^{-1}ds= 2\int_{s_{0}}^{s_{1}}\frac{g(s)}{f(s)}\frac{f(s_{0})}{\sqrt{f^{2}(s)-f^{2}(s_{0})}}ds \hspace{.2cm},
\end{eqnarray}
where $s_{0}$ and $s_{1}$ are, respectively, de equilibrium position of the bottom of the string and the position of its endpoints. 

Since the action has dimensions of Energy $\times $ time, the energy of the configuration associated with the string will be given by \eqref{eqn6}
\begin{eqnarray}\label{eqn8}
E = \frac{1}{\pi\alpha'} \int_{s_{0}}^{s_{1}}\frac{g(s)}{f(s)}\frac{f^{2}(s)}{\sqrt{f^{2}(s)-f^{2}(s_{0})}}ds \;.
\end{eqnarray}

Performing the change of variable $s= \frac{R^{2}}{z}$, where $R$ is the AdS radius, we have  $z_{0} = \frac{R^{2}}{s_{0}} $, $  z_{1} = \frac{R^{2}}{s_{1}}$, and $ds=-\frac{R^{2}}{z^{2}}dz$. 
We take the limit $s_{1} \to \infty $, which means that $z_{1} \to 0$ and we can rewrite \eqref{eqn7} and \eqref{eqn8} as:
\begin{eqnarray}
l= 2\int_{0}^{z_0}\frac{g(z)}{f(z)}\frac{f(z_{0})}{\sqrt{f^{2}(z)-f^{2}(z_{0})}}\frac{R^{2}}{z^{2}}dz \label{eqn9} \hspace{.2cm} \\
E = \frac{1}{\pi\alpha'} \int_{0}^{z_0}\frac{g(z)}{f(z)}\frac{f^{2}(z)}{\sqrt{f^{2}(z)-f^{2}(z_{0})}}\frac{R^{2}}{z^{2}}dz \label{eqn10} \hspace{.1cm} .
\end{eqnarray}
Using the metric \eqref{metric}
we have $ f(z) =  h(z)\frac{R^{2}}{z^{2}} $ and  $g(s)  = h(z) $. 
Then, from Eqs. \eqref{eqn9} and \eqref{eqn10} one gets \eqref{e} and \eqref{l}.

\end{document}